\documentclass[notitlepage,nofootinbib,amsmath,longbibliography,amssymb,twocolumn,groupedaddress]{revtex4-1}
\usepackage{amsmath,amssymb,graphicx}
\usepackage{epsf}
\usepackage{epstopdf}
\usepackage{mathtools}
\usepackage {amssymb}
\usepackage{mathcomp}
\usepackage{cancel}
\usepackage{textcomp}
\usepackage{enumitem}
\usepackage{physics}
\usepackage{color}
\newcommand{\nc}{\newcommand}
\nc{\ba}{\begin{eqnarray}}
\nc{\ea}{\end{eqnarray}}
\newcommand\be{\begin{equation}}
\newcommand\ee{\end{equation}}

\newcommand{\calF}{{\cal{F}}}

\newcommand{\calV}{{\mathcal{V}}_{\text{int}}}
\newcommand{\calW}{{\mathcal{W}}_{\text{int}}}
\nc{\x}{{\bf{x}}}

\usepackage[colorlinks=true, linkcolor=blue, citecolor=magenta, hyperfootnotes=true]{hyperref}

\begin{document}
\title{A No-Go Theorem for the Mass-Radius Relation of Solitons}
\author{Mohammad Hossein Namjoo}
\email{mh.namjoo@ipm.ir}
\affiliation{School of Astronomy, Institute for Research in Fundamental Sciences (IPM),\\ Tehran, Iran, P.O. Box 19395-5531}

\begin{abstract}

We prove a no-go theorem for the mass-radius relation of localized and stable field configurations, known as solitons. Defining the mass-radius index by $\Gamma \equiv \frac{{\rm{d}}\ln M}{{\rm{d}}\ln R}$, for real scalar field theories in $d$ spatial dimensions, we show that typical non-topological, non-relativistic, and spherically symmetric solitons cannot have $\Gamma$ in the range $[0, d]$. The forces considered originate from gradient energy, self-interaction, and gravitation, with the typicality assumption excluding the fine-tuned region of the parameter space where all three forces have comparable strength. Importantly, the theorem works for an arbitrary self-interaction that, in the relativistic theory, is allowed to be  non-power-law in the field, be non-analytic around the classical vacuum (where the field amplitude vanishes), or to include derivative couplings. Additionally, the theorem makes no assumptions about the explicit form of the soliton's density profile or the behavior of $\Gamma$ as a function of $R$. We also argue that the same exclusion applies to compact objects formed from self-gravitating, non-relativistic, barotropic fluids with arbitrary equations of state. As a consequence for cosmology, it is worth noting that observations favor a core in the centers of dark matter halos with $\Gamma \simeq 1.7$, which (for $d=3$) lies approximately in the middle of the excluded range. Therefore, proposals such as ultra-light or fluid-like dark matter models are essentially ruled out as natural explanations for halo cores, provided other astrophysical effects are negligible. 
\end{abstract}
\maketitle
\section{Introduction}
\label{Sec:intro}
Studying the various properties of localized and stable field configurations, known as solitons,\footnote{Depending on the context and the forces involved, different names are used in the literature for localized, stationary, and stable solutions of a field theory. In this paper, we refer to all of them as ``solitons".} is of interest in many areas of physics \cite{DauxoisPeyrard2010, Gu1995}. {A property of solitons that has been extensively studied in physics literature is their mass-radius relation. This study provides insights into the forces involved in soliton formation \cite{Schiappacasse:2017ham}, the region in the parameter space that allows the formation of celestial objects \cite{Chandrasekhar:1931ih,Mielke:1997re} and the physics governing Bose-Einstein condensation \cite{Chavanis:2011zi}.}

A notable application of solitons is their potential to address the so-called core-cusp problem in cosmology. Observations of galactic rotation curves suggest the presence of a core at the center of dark matter (DM) halos. This contrasts with the standard DM scenario, where simulations show no evidence of a core; instead, the density profile appears cuspy. On the other hand, for some other DM candidates, such as ultra-light  \cite{Guth:2014hsa} or fluid (or superfluid) DM \cite{Berezhiani:2025maf} scenarios, it is natural to expect the formation of a solitonic core due to different governing dynamics, which could explain the observed phenomena \cite{Marsh:2015wka,Hui:2016ltb,Goodman:2000tg,DelPopolo:2021bom}.

\begin{figure}
	\includegraphics[scale=.7]{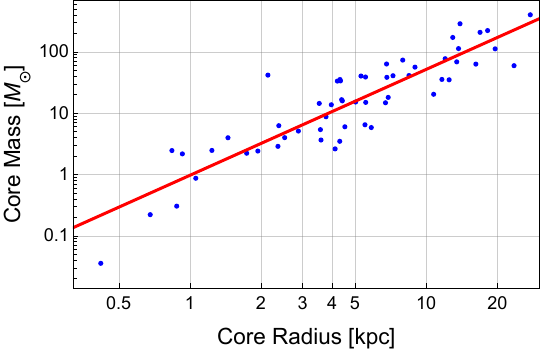}
	\caption{The mass-radius diagram for the core of the dark matter halos. The blue dots represent the observational data reported in Ref.~\cite{Rodrigues:2017vto} and the red line is the best fit proportional to $R^{1.7}$. }
	\label{core}
\end{figure}

 Studies of the DM core show that there is indeed a correlation between their mass and radius.\footnote{{It is interesting to note that there are also indications of a correlations between the core and the halo mass for the ultra-light dark matter models \cite{Schive:2014hza,Chan:2021bja,Mocz:2017wlg,Nori:2020jzx,Zagorac:2022xic}. We will not discuss this phenomenon in this paper.}} In Fig.~\ref{core}, we display how the core mass varies with its radius based on the data from Ref.~\cite{Rodrigues:2017vto}. Despite some scatter, a clear correlation between core mass and radius is evident. Assuming a power-law relationship, the best fit is $M \propto R^{1.7}$ \cite{Deng:2018jjz}.\footnote{{ Note that Ref.~\cite{Rodrigues:2017vto} uses the Burkert profile to identify the core. Interestingly, Ref.~\cite{Banares-Hernandez:2023axy} uses a different dataset along with the soliton+NFW profile but finds a consistent result; the authors' best-fit is $M \propto R^{1.8}$. A more general analysis of how the core properties depend on the choice of profile is a valuable direction for further research. }} Any DM-only scenario that ought to solve the core-cusp problem must, in addition to predicting a core, also be consistent with the observed mass-radius relation.

In this paper, we establish a no-go theorem for the mass-radius relation of solitons. Since our goal is to prove the theorem in the most general form, we do not restrict the analysis to a power-law relation. Instead, we work with the mass-radius index, which we define as
\ba 
\label{Gamma_def}
\Gamma \equiv \frac{d\ln M}{d\ln R}.
\ea 
We then show that, for typical non-topological, non-relativistic, and spherically symmetric solitons in $d$ spatial dimensions, the region $[0,d]$ is excluded for $\Gamma$. We will be more explicit about the meaning of the typicality assumption in the course of the proof, but it aims at excluding fine-tuned situations. Notably, the theorem applies to any self-interaction and does not rely on a specific soliton profile. We also argue that the same no-go theorem holds for a barotropic fluid when it forms a stable, spherically symmetric object balancing Newtonian self-gravity. It is notable that, in the case of $d=3$, this result rules out DM solitons as a natural explanation for the halo core, as long as the soliton’s self-forces are assumed to be the only relevant sources for the dynamics of the system {(that is, as long as other effects such as baryonic feedback or the influence of supermassive black holes are negligible). }

{The disagreement between the observational data and the ultra-light DM models has already been discussed in Ref.~\cite{Deng:2018jjz}, but under more restrictive assumptions than are needed for the theorem that we will present in this paper. Explicitly, in Ref.~\cite{Deng:2018jjz}, three spatial dimensions, a power-law potential, a Gaussian profile and a power-law mass-radius relation are assumed.  We relax all these assumptions. Additionally, the actual excluded region is made explicit in this work. Therefore, we generalize the results of Ref.~\cite{Deng:2018jjz} in various ways, as is evident from the statement of the theorem.}

In Sec.~\ref{Sec:NREFT}, we discuss the non-relativistic effective field theory (NREFT) as a useful tool for studying non-relativistic solitons. In Sec.~\ref{Sec:energies}, we present energy functionals based on the NREFT that will be used in Sec.~\ref{Sec:proof} for the proof of the no-go theorem. In \ref{Sec:fluid} we extend the proof to systems involving a barotropic fluid. We conclude the paper in Sec.~\ref{Sec:conclusion} mainly with a few remarks on possible future research directions.

\section{Non-relativistic effective field theory}
\label{Sec:NREFT}
In this section, we briefly discuss a general form of the Lagrangian for the leading order NREFT. To be as general as possible, we mainly focus on an NREFT in the bottom-up approach. However, we start with a brief discussion of the top-down approach to highlight the generality of the analysis from a different perspective. Consider a relativistic, real, massive scalar field theory with a generic self-interaction in arbitrary spacetime dimensions as:
\begin{equation}
	\label{Lphi}
	\mathcal{L}_{\phi} = X - \dfrac{1}{2} m^2 \phi^2 - V_{\text{int}} (X,\phi)\, ,
\end{equation}
where we defined 
$X \equiv  - \frac{1}{2} \eta^{\mu\nu} \partial_{\mu}\phi \partial_{\nu}\phi \, $,
using the mostly positive metric signature. For a consistent non-relativistic limit, we assume that the mass term dominates over the self-interaction potential. The self-interaction term is arbitrary and is allowed to be non-power law, non-analytic around the classical vacuum ($\phi=0$), or to include derivative coupling. Examples of each class are dilatons/axions, potentials with logarithmic radiative corrections, and the Dirac-Born-Infeld theory, respectively. (Therefore, considering such a general potential is well-motivated theoretically.) Since the relativistic Lagrangian must be Lorentz invariant, the derivative coupling is restricted to be a function of $X$. 

For deriving the NREFT in the large mass limit of the  Lagrangian Eq.~\eqref{Lphi}, it is convenient to introduce a new complex scalar field $\psi$
via
\begin{align}
	\label{full-redef}
	&\phi = \dfrac{1}{\sqrt{2m}} \left( \psi e^{-imt} +\text{c.c} \right), \,
	&\dot \phi = -i \sqrt{\dfrac{m}{2}} \left( \psi e^{-imt} -\text{c.c} \right).
\end{align}
This field redefinition separates the fast and slow dynamics of the system and, in the limit of a free theory, transforms the Klein-Gordon equation for $\phi$ to the Schr\"odinger equation for $\psi$. The Lagrangian for $\psi$, which is a slowly varying field, will contain terms that oscillate rapidly with frequencies of $m$ or higher, which average to zero and can be ignored, to leading order in the NREFT. A systematic derivation of the NREFT, accounting for the backreaction of oscillatory terms on the dynamics of the slowly varying field, can be found in Ref.~\cite{Namjoo:2017nia}, and the analysis for arbitrary potentials is presented in Ref.~\cite{Modirzadeh:2025gjd}. What matters to us in this paper is to note that the resulting NREFT will have a global $U(1)$ symmetry with the conserved charge $N=\int |\psi|^2 \dd[d]x$, which is the (conserved) number of particles in the non-relativistic limit. Therefore, very generally, the leading order NREFT takes the following form
\begin{equation}
	\mathcal{L}_{\psi,\psi^*} = \dfrac{i}{2}(\psi^*\dot{\psi}-\psi\dot{\psi^*})- \dfrac{1}{2m} \grad\psi .\grad\psi^* 
	- \mathcal{V}_{\text{int}}(|\psi|^2) +...
	\label{Lag-psis}
\end{equation}
where terms that are neglected are suppressed by higher derivatives, while all terms suppressed by higher powers of coupling constants can be incorporated into the general self-interaction potential $\calV$. Although the self-interaction in the relativistic theory $V_{\text{int}}$ is allowed to contain derivatives, $\calV$ is only a function of $|\psi|^2$ (not its derivative). This is a consequence of the NREFT, which captures low-energy dynamics where derivatives cause additional suppression and appear at higher orders. To see how an $X$-dependent self-interaction can lead to a potential in the NREFT that has no derivatives, note that at the leading order in the low energy limit, the spatial derivatives can be neglected, thus we have
$X \simeq \frac{1}{2} \dot \phi^2$.
By using Eq.~\eqref{full-redef}, we can replace $\dot \phi$ with the non-relativistic fields $\psi$ and $\psi^*$. Neglecting oscillatory terms, this yields $X \simeq \frac{m}{2} |\psi|^2$, which has no derivatives and can be used to substitute for $X$ in the self-interaction term. As mentioned earlier, we do not assume any specific form for $V_{\text{int}}$ (in the top-down NREFT), nor for $\calV$ (in the bottom-up NREFT).  The only restriction is the validity of the NREFT, which requires the $U(1)$ symmetry to ensure conservation of particle number in the non-relativistic limit, and that higher derivative terms are more suppressed and can therefore be neglected.\footnote{It is conceivable that, in a fine-tuned situation, a precise cancellation results in an NREFT with a leading order self-interaction involving derivatives. However, first, the no-go theorem is claimed to apply to typical solitons, thus excluding this unrealistic scenario. Second, it can be shown, using a similar method that follows (Sec.~\ref{Sec:proof}), that even a generic potential of the form $|\psi|^{2 p}\, (\nabla \psi.\nabla \psi^*)^{q}$ as the leading self-interaction cannot violate the no-go theorem for arbitrary constants $p \geq 0$ and $q \geq 1$.}

\section{Energy functionals}
\label{Sec:energies}
In this section, we present the expression for the total energy of a stationary system of the non-relativistic field. Including gravity, it consists of four terms 
\begin{equation}
	\label{Hamiltonian}
	H_{\text{T}} = H_{\text{m}}  + H_{\text{grad}} + H_{\text{int}}  + H_{\text{grav}} \, ,
\end{equation}
where $H_{\text{m}},  H_{\text{grad}} , H_{\text{int}}$, and $H_{\text{grav}}$ are the rest mass energy, gradient energy,  self-interaction energy, and gravitational energy,  respectively, and are given by
\ba 	
& H_{\text{m}} \equiv m \int \dd[d]{r} \abs{\psi}^2 \ , \quad 
H_{\text{grad}} \equiv \dfrac{1}{2m} \int \dd[d]{r} \grad{\psi} \vdot \grad{\psi^*} \ , \nonumber \\
&H_{\text{int}} \equiv \int \dd[d]{r} \mathcal{V}_{\text{int}}  \, , \\ \nonumber 
	\label{energies}
&	H_{\text{grav}}  \equiv  -\dfrac{G_d\, m^2}{2} \int \dd[d]{r} \int \dd[d]{r'} \dfrac{\abs{\psi({\bf r})}^2 \abs{\psi({\bf r'})}^2}{|{\bf r}-{\bf r'}|^{d-2}}  \, .
\ea 
The rest mass $H_{\text{m}}$ is the main contribution to the total energy and determines the mass of the compact object, but it does not affect the dynamics of the non-relativistic system. $H_{\text{grad}}$ and $H_{\text{int}}$ can be derived from the Lagrangian  Eq.~\eqref{Lag-psis} through a Legendre transformation. $H_{\text{grav}}$ applies when $d> 2$, where $G_d$ is the gravitational constant in $d$ dimensions. We study the effect of gravitational energy for $d> 2$ first and then point out how the same conclusion can be reached for $d=2$. It is known that gravity cannot form stable objects in $d=1$, so the no-go theorem applies in all dimensions. 

From now on, we proceed under the assumption of spherical symmetry. The standard procedure is then to assume a specific profile for the soliton to evaluate the energy functionals in Eq.~\eqref{energies}. However, to stay as general as possible, we do not make any assumptions about the profile (nor do we need to) and instead express the total energy of the system as 
\ba 
\label{energy}
E = N m +c_1 \dfrac{N}{mR^2} -c_2 \dfrac{G_d \, m^2 N^2}{ R^{d-2}} + s\,  R^d \, \calW(N/R^d)\, , \nonumber \\
\ea 
where $R$ is the size of the object, and $s=2 \pi^{\frac{d}{2}}/\tilde \Gamma(\frac{d}{2})$ is the surface area of a ($d-$1)-sphere, with $\tilde \Gamma(.)$ being the Gamma function. $c_1$ and $c_2$ are two constants, whose values depend on the profile. We only require them to be positive, which is determined by the fact that the gradient energy causes a repulsive force, while Newtonian gravity is attractive. For the first three terms, the appropriate factors of $N$ and $m$ can be observed from Eq.~\eqref{energies} (and recall that $N=\int |\psi|^2 \, \dd^d r$). The powers of $R$ can also be derived through dimensional analysis. For the last term, besides the spherical symmetry, we used the fact that the potential is only a function of $|\psi|^2$, and defined 
\ba 
\label{calW_def}
\calW  \equiv \dfrac{1}{R^d} \int_0^{\infty} \calV \big(|\psi(r)|^2 \big)\,  r^{d-1} \dd  r\, .
\ea 
Again, a dimensional analysis reveals that since $|\psi|^2$ is the number density, $\calW$ has to be a function of $(N/R^d)$, which is made explicit in Eq.~\eqref{energy} and will be used in the proof of the no-go theorem. To explicitly see how this dependence arises as well as the explicit form of $c_1$ and $c_2$ for a generic profile of the form $\psi(r) \propto \sqrt{ \frac{N}{ R^d}}\, e^{-(\frac{r}{R})^\delta}$ (for $d=3$ but with arbitrary positive $\delta$), see Ref.~\cite{Modirzadeh:2025gjd}.

A localized object can form if, for fixed $N$, the energy reaches an extremum as a function of $R$. That is, we require $\pdv{E}{R}=0$. This must be accompanied by the stability condition $\pdv[2]{E}{R}>0$. In the next section, we will use these two conditions to prove the theorem when two forces are at work (and the third is negligible) --- which is expected to be valid for typical solitons. 

\section{Proof of the no-go theorem}
\label{Sec:proof}
We now turn to the main task of this paper. We start with studying the mass-radius relation when self-interaction balances either the gradient energy or gravity, requiring the self-interaction to be attractive or repulsive, respectively. To analyze both scenarios within a single framework, we consider the total energy in the following form
\ba 
\hat E \equiv \dfrac{E }{s} = \alpha N^{\beta} R^{\gamma}+ R^d\,  \calW(\hat n)\, ,
\ea 
where we defined $\hat E=E/s$ and $\hat n=N/R^d$ for convenience, and $\alpha$, $\beta$ and $\gamma$ are three constants that take different values for different forces. Explicitly, for the triplet $(\alpha,\beta,\gamma)$, we have $(\frac{c_1}{s m}, 1,-2)$ for the gradient energy and $(-\frac{c_2 G_d m^2}{s }, 2, 2-d)$ for gravity. We will see that the only requirement for the triplet, needed for the proof of  the no-go theorem is that 
\ba 
\label{con_params}
\alpha \, \gamma \, (\gamma +d\, \beta -d) >0\, ,
\ea 
which is satisfied for both the gravitational and gradient energies. Note that this constraint may be satisfied by many other forces, suggesting that the theorem is more general than the claim of this paper. 

For the existence of  the stationary objects we require 
\ba 
\label{E'}
\pdv{\hat E}{R} = 
 \alpha \gamma N^{\beta}  R^{\gamma-1} + d\, R^{d-1} \calF = 0 \, ,
\ea
where, recall that, the partial derivative holds $N$ constant, and we defined 
$
\calF \equiv \calW - \hat n \calW'\, ,
$
where the prime denotes the derivative with respect to the argument. For this algebraic equation to have real solutions, we need
\ba 
\label{con_w}
\alpha \gamma \calF  <0.
\ea 
Presumably, there are additional constraints for the existence of the solution, but this is the only constraint we need. By changing the number of particles $N$ the field configuration will be organized with the appropriate $R$ that satisfies Eq.~\eqref{E'} with the new value of $N$. Therefore, thinking of $R$ and $\hat n$ as two independent variables, differentiating Eq.~\eqref{E'} yields
\ba 
\label{diff}
(\gamma+d\, \beta -d)\,  \text{d} (\ln R) =-\bigg(\beta + \dfrac{\hat n^2 \calW''}{\calF}\bigg)\, \text{d}(\ln \hat n ).
\ea 
The mass of the object is $M=N m=m \hat n R^d$, thus $\Gamma=d+ \dv{\ln  \hat n }{\ln R} \, $, and from Eq.~\eqref{diff} we have
\ba 
\label{Gamma_nhat}
\Gamma=d- \dfrac{(\gamma+d\, \beta -d) \calF}{\beta \calF +\hat n^2 \calW''}.  
\ea 
The stability of the solution requires 
\ba 
\label{E''}
d^{-1} R^{2-d} \, \pdv[2]{\hat E}{R} = (d-\gamma) \calF +d\,  \hat n^2 \, \calW'' >0\, ,
\ea 
where we also used Eq.~\eqref{E'} for some simplification. 
Using Eq.~\eqref{Gamma_nhat} to substitute  $\calW''$ in Eq.~\eqref{E''} results in 
\ba 
\label{con_Gamma_interm}
\big(\dfrac{\Gamma}{\Gamma-d}\big)   (\gamma+d\, \beta -d)  \calF <0. 
\ea 
Finally, using the inequalities Eq.~\eqref{con_params} (holds for the gradient energy and gravity), and Eq.~\eqref{con_w} (required for the existence of the solution), Eq.~\eqref{con_Gamma_interm} (the stability condition) reduces to $\frac{\Gamma}{\Gamma-d}>0$, or
\ba 
\label{Gamma_constraint}
\Gamma>d \quad \text{or} \quad \Gamma <0 \, .
\ea 
In other words, $\Gamma$ cannot be in the range $[0,d]$, as claimed.  

Another branch of typical solitons can form when self-interaction is weak and the two other forces balance each other. In this case, a similar analysis for the situation where only the second and the third terms of Eq.~\eqref{energy} are relevant reveals that $\Gamma=d-4$. Naively, this might suggest that it is possible to violate the no-go theorem for $d>4$. However, it is easy to see that the stability condition ($\pdv[2]{\hat E}{R}>0$) requires $d<4$. Therefore, interestingly, the no-go-violating solutions cannot be stable.\footnote{Having stable objects in $d\geq 4$ is known to be harder than $d=3$, making $d=3$ anthropically preferred since it allows for the existence of compact objects such as stars, necessary for life  \cite{Chavanis:2006pf,bechhoefer1993}.} 

For $d=2$, the gravitational potential is logarithmic, and the total gravitational energy behaves like $E_{\text{grav}} \sim G_2 \, m^2 N^2 \ln R$. It is straightforward to repeat the analysis for this energy, from which one finds that the interval $[0,2]$ is excluded for $\Gamma$. This concludes the proof of the no-go theorem in arbitrary spacetime dimensions. 

Note that there is a small window in the parameter space where all three forces are roughly equal in size and contribute to soliton formation. The no-go theorem is not claimed to apply in this fine-tuned (and therefore unrealistic) situation — see Ref.~\cite{Chavanis:2011zi} for an analysis of this situation in a specific example. Even if this region is physically acceptable, it cannot explain phenomena such as the mass-radius relation of DM halo cores, since they require maintaining the appropriate mass-radius relation over at least a few orders of magnitude in radius. 

\section{Barotropic fluids }
\label{Sec:fluid}
Another class of stable objects that has been widely discussed consists of a fluid, the pressure of which balances another force, such as gravity. Here, we briefly argue that a barotropic fluid with an arbitrary equation of state, when encounters Newtonian gravity, has the same exclusion region in the mass-radius relation. Denote the barotropic fluid's energy density and pressure by $\rho$ and $P=P(\rho)$, respectively. Working with the scaling behavior of different variables, we have $\rho \sim M/ R^d$. The pressure creates a repulsive force with the scaling $F_P \sim P(\rho) R^2$ which balances the attractive gravitational force which scales like $F_G \sim -G_d M^2 R^{1-d}$. Therefore, the mass-radius index is given by 
\ba 
\label{Gamma_fluid}
\Gamma = d +\dfrac{d-1}{\kappa-2}\, ,
\ea 
where $\kappa \equiv \dv{\ln P}{\ln \rho}$ which is known as the adiabatic index. The stability condition can be found by  taking the derivative of the net force with respect to the radius in a constant mass.\footnote{More explicitly, since ${\bf F} \sim -{\bf \nabla} E$, where $\bf F$ is the net force, $\pdv[2]{E}{R}>0$ implies $\pdv{F}{R}<0$.} This results in the requirement $\kappa > 1+\frac1d$, which is a well-known stability requirement, at least for $d=3$ \cite{ShapiroTeukolsky1983}. Converting this constraint to a constraint for $\Gamma$ using Eq.~\eqref{Gamma_fluid} we obtain exactly the allowed region as given by Eq.~\eqref{Gamma_constraint}.

Remarkably, we see that the same exclusion in $\Gamma$ occurs for this scenario. However, given the findings of Sec.~\ref{Sec:proof}, this may not be surprising. To see this, note that the NREFT as described by the Lagrangian Eq.~\eqref{Lag-psis} may be interpreted as a fluid, for which the energy density is $\rho \simeq m |\psi|^2$. The fluid's pressure has two contributions. The so-called ``quantum pressure," which accounts for the gradient energy discussed in Sec.~\ref{Sec:energies} and the classical pressure resulting from self-interaction. Neglecting the former, the latter contribution is given by (see Ref.~\cite{Modirzadeh:2025gjd} for a rigorous derivation):
\ba 
\label{Peff}
P_{\text{eff}} = \abs{\psi}^2   {\mathcal{V}}_{\text{int},|\psi|^2}  - {\mathcal{V}}_{\text{int}}\, ,
\ea 
where ${\mathcal{V}}_{\text{int},|\psi|^2}$ is the derivative of the potential with respect to $|\psi|^2$. Since $\rho \simeq m |\psi|^2$ and $\calV =\calV(|\psi|^2)$, the right-hand side of Eq.~\eqref{Peff} can be expressed in terms of the energy density. Therefore, a fluid with a given equation of state can be described by a non-relativistic field with an appropriate self-interaction. To further explore the analogy, which also supports the applicability of the mass-radius relation of Sec.~\ref{Sec:proof} to the fluid scenario,  use the first law of thermodynamics for an adiabatic process $du =\frac{P}{\rho^2} d\rho$, where $u$ is the specific internal energy. Replace the quantities on the right-hand side by their field counterparts and integrate. This yields $u = \calV/\rho$. The total internal energy, which we denote by $U$, is then given by
\ba 
U =\int u \, \rho \, \dd[d] r = \int \calV \, \dd[d] r\, ,
\ea 
which is precisely $H_{\text{int}}$ according to Eq.~\eqref{energies}. Therefore the total internal energy of a fluid arising from an arbitrary equation of state corresponds to a total self-interaction energy of a non-relativistic scalar field. This duality  implies the applicability of the  no-go theorem in both situations.

\section{Conclusion}
\label{Sec:conclusion}
We have shown that the mass-radius index of solitons --- defined by Eq.~\eqref{Gamma_def} --- cannot lie within the range $[0,d]$, where $d$ is the number of spatial dimensions. This no-go theorem applies to typical non-topological, non-relativistic, and spherically symmetric solitons formed from a real scalar field. The typicality assumption excludes fine-tuned situations where all three forces (gravity, self-interaction, gradient) are of comparable size, or where precise cancellations make derivatively suppressed self-interactions in the non-relativistic effective field theory (NREFT) the dominant contribution. Despite these restrictions, the theorem is very general; it applies to any spatial dimension, any self-interaction that may appear in the relativistic scalar field theory, and makes no assumptions about the soliton’s density profile. Its broad applicability is achieved thanks to the NREFT, whose symmetries are restrictive enough to allow a model-independent analysis. We have also argued that the same no-go theorem applies to a barotropic fluid with an arbitrary equation of state when it is in equilibrium with its own gravity.

It would be valuable to investigate extending the no-go theorem beyond the assumptions made here. Examining the theorem’s applicability to other scenarios, such as complex scalar fields or multiple field setups, would also be worthwhile. {The validity of the theorem in more complex --- and more realistic --- situations such as when baryonic feedback \cite{Chan:2021bja} or the gravitational effect of supermassive black holes \cite{Chavanis:2019bnu} are taken into account is worth exploring.}
While we have briefly discussed the consequences of the no-go theorem for cosmology, its implications for other areas of physics remain to be explored. Finally, a deeper understanding of the physics behind the no-go theorem, beyond the insights provided in this paper using the NREFT techniques, remains an intriguing open problem. 

%
{}

\end{document}